\documentclass[epj,twocolumn]{webofc}
\usepackage[varg]{txfonts}   
\woctitle{Powders \& Grains 2017}
\begin{document}
\title{Dissipative lateral walls are sufficient to trigger convection in vibrated granular gases}
%
%

\author{  \firstname{Giorgio} \lastname{Pontuale}\inst{1}\fnsep\thanks{\email{giorgio.pontuale@isc.cnr.it}} \and
          \firstname{Andrea} \lastname{Gnoli}\inst{1}\fnsep\thanks{\email{andrea.gnoli@isc.cnr.it}} \and
          \firstname{Francisco} \lastname{Vega Reyes}\inst{2}\fnsep\thanks{\email{fvega@unex.es}} \and
          \firstname{Andrea} \lastname{Puglisi}\inst{1}\fnsep\thanks{\email{andrea.puglisi@roma1.infn.it}} 
}

\institute{Istituto dei Sistemi Complessi—CNR and Dipartimento di Fisica, Università di Roma Sapienza, Piazzale Aldo Moro 2, 00185 Rome, Italy
\and
Departamento de Física and Instituto de Computación Científica Avanzada (ICCAEx), Universidad de Extremadura, 06071 Badajoz, Spain
          }

\abstract{%
  Buoyancy-driven (thermal) convection in dilute granular
  media, fluidized by a vibrating base, is known to appear without the
  need of lateral boundaries in a restricted region of parameters
  (inelasticity, gravity, intensity of energy injection). We have
  recently discovered a second buoyancy-driven convection effect which
  occurs at any value of the parameters, provided that the impact of
  particles with the lateral walls is inelastic (Pontuale et al., Phys. Rev. Lett. 117, 098006 (2016)). It is understood that this
  novel convection effect is strictly correlated to the existence of
  perpendicular energy fluxes: a vertical one, induced by both bulk
  and wall inelasticity, and a horizontal one, induced only by
  dissipation at the walls. Here we first review
  those previous results, and then present new experimental and
  numerical data concerning the variations of box geometry, intensity
  of energy injection, number of particles and width of the box. }
\maketitle
%
%
\section{Convection in granular media, from slow dense configurations to dilute gas-like setups}
\label{intro}
Vibration of granular materials frequently leads to convective
patterns~\cite{AT06}. However, the particular mechanism for sustaining and
control of a convective dynamics depends upon the particular granular
phase under scrutiny~\cite{JNB96b}. At high packing
fraction~\cite{andreotti13} and low fluidization, ``dense convection''
has been first identified in the
90's~\cite{laroche89,knight93,ehrichs95,knight96} and explained
through an asymmetric tangential friction at the lateral
walls~\cite{gallas92,knight97,grossman97b} or the formation of
unstable heaps at the free surface~\cite{aoki96}. In granular
gases~\cite{D01,poeschel,puglio15}, on the contrary, the first
discovered mechanism is bulk buoyancy-driven thermal convection
(BBD-TC), initially observed in
simulations~\cite{ramirez00,sunthar01}. As in molecular
liquids~\cite{landau71,chandra81}, it is driven by the buoyancy force
associated to temperature/density gradients: in BBD-TC such gradients
appear spontaneously because of the bulk inelasticity of grain-grain
collisions~\cite{VU09}. This is a bulk effect: indeed, no particular conditions on the lateral or top
boundaries are required for its appearance (for instance
in~\cite{ramirez00} and~\cite{paolotti04} it appears also with elastic lateral walls) and in
fact an analytical study of the stability of the granular gas
hydrostatic state {\em without lateral walls} (lateral periodic
boundary conditions) has shown the emergence of BBD-TC in a region of
system's parameters~\cite{meerson1,meerson2}. When lateral walls are
present (elastic or inelastic) a downward flow velocity is always
observed near lateral walls, perhaps because of a reduced buoyancy
originated from enhanced dissipation~\cite{ramirez00,talbot02}. After
the first experimental evidence in 3D~\cite{wildman01}, other
experiments in quasi-2D setups have demonstrated the presence of
convection in granular gases~\cite{lohse07}, but with contradicting features: some of
them agree quantitatively with analytical theories where lateral walls are
absent~\cite{lohse10}, while others~\cite{windows13} strongly indicate that  convection is almost completely
killed when wall inelasticity goes to zero, a scenario - incompatible
with BBD-TC theories~\cite{meerson1,meerson2} (see~\cite{talbot02} for numerical simulations with similar results).

Recently we have demonstrated the existence of a convection mechanism
for granular gases which is alternative to BBD-TC and which
necessarily appears in all granular gases fluidized by a vibrating
base under gravity, with the only necessary condition being the presence of inelastic lateral walls~\cite{noi}. Such a
mechanism has been called dissipative-lateral-wall-induced thermal convection, ``DLW-TC''. In this recent study we have employed experiments
and simulations to isolate DLW-TC from BBD-TC. In Section 2 we review these
recent results, while in Section 3 we present  new experimental
and numerical data. Conclusions are drawn in Section 4.
\section{DLW-TC}
\label{continuity}
\begin{figure}[tb!]
  \centering
\includegraphics[width=7.5cm,clip]{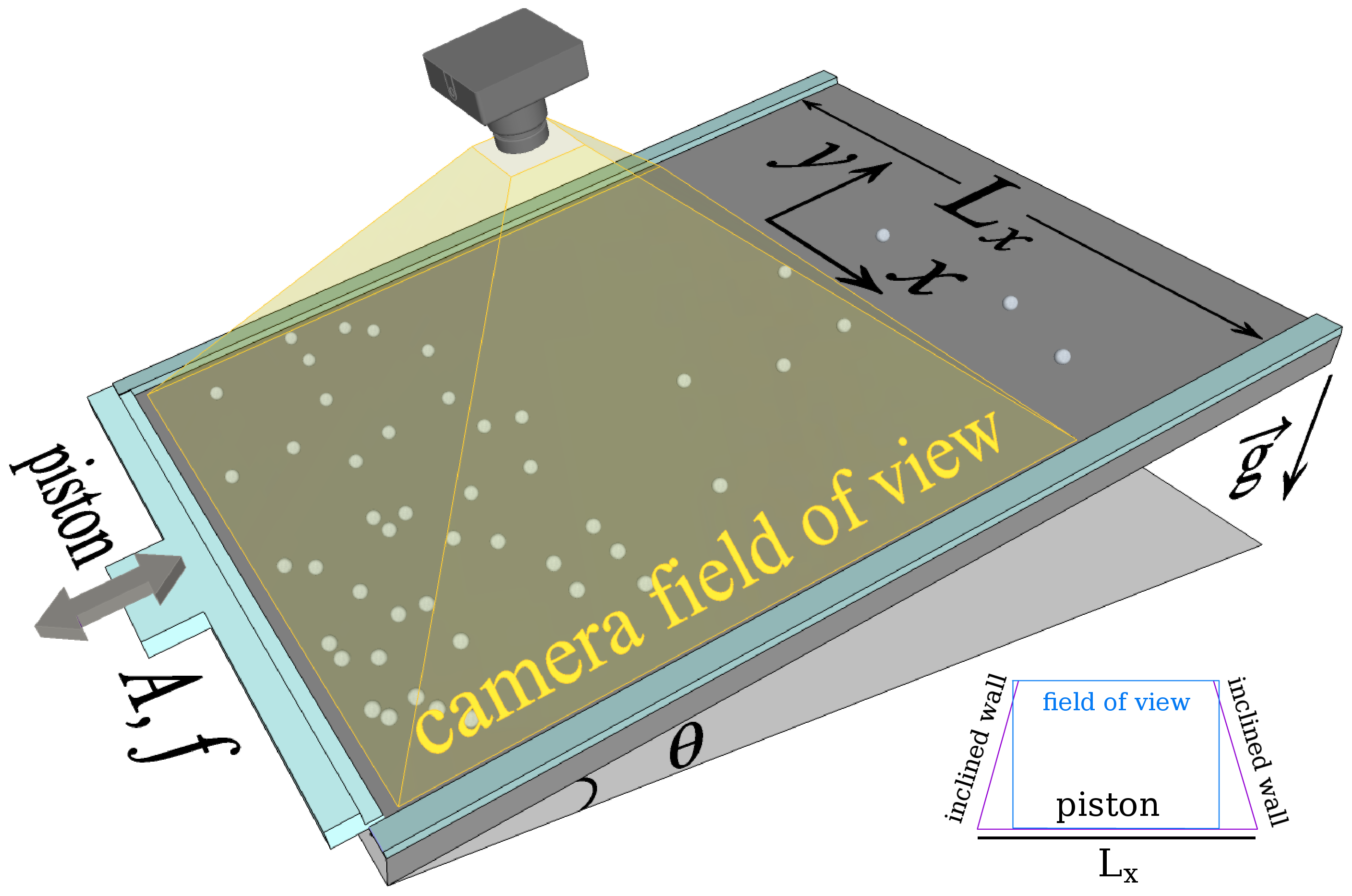}
\caption{Sketch of the experimental setup. The real length (along $y$) of the inclined plane, $L_y \gg L_x$, is not marked as it is not essential in the description of the system. In the lower-right
  corner a sketch of the second setup with inclined lateral walls is
  shown, with a representation (blue square) of the restricted field
  of view for the fast camera. \label{fig:setup}}
\end{figure}
Let us consider (similarly to the setup sketched in Fig.~\ref{fig:setup}) a 2D low density
gas (easily generalized to 3D systems) of identical solid
disks of mass $m$ confined by two inelastic parallel
walls. The bottom wall provides
energy to the system, for instance through steady vibration or (in
numerics) in the form of a thermostat. The presence of a top wall or open top boundary does not change our
reasoning.  A constant gravity field $g$ acts downwards along
the vertical ($y$) direction. In the dilute limit $p=nT$ (with $p$ the pressure, $n$ the number
density, and $T$ the granular temperature)~\cite{D01}. An outgoing energy flux is always originated
at a dissipative wall~\cite{N99}, yielding in our case $\partial T /\partial x \neq 0$ at the lateral walls.
Momentum balance in the absence of macroscopic flow (hydrostatic) states that
\begin{subequations}
\label{2d_balance}
\begin{align}
\partial_x p=\partial_x(n(x,y)T(x,y))=0\\
\partial_y p=\partial_y (n(x,y)T(x,y))=-mgn(x,y).
\end{align}
\end{subequations}
The first equation gives $p(x,y) \equiv p(y)$, which, used in the second equation,
sets $n(x,y) \equiv n(y)$ and as a consequence $T(x,y)\equiv T(y)$. This is in contradiction with the horizontal
temperature gradient required by dissipation at the lateral walls. Convection must be present.

In~\cite{noi} we have discussed an
experiment (inspired on previous works~\cite{kudrolli97b,kudrolli00}) under low-gravity conditions
(see Fig.~\ref{fig:setup} and~\cite{noi}, including Supplemental Materials, for details) with $N$
spherical steel beads (diameter $d=1~\mathrm{mm}$) moving inside a
cuboid of dimensions $L_x=175~\mathrm{mm}$, $L_y=600~\mathrm{mm}$ and
$L_z=1.5~\mathrm{mm} = 1.5 d$, thus assuring for a quasi-2D dynamics
restricted to the $xy$ plane. A small tilt angle $\theta$ with respect
to the horizontal guarantees an effective low value for gravity
$g_{eff} \approx (5/7) g \sin(\theta)$, where $g$ is Earth's gravity
acceleration~\cite{blair03}.  The two lateral ``walls'' are made of
polycarbonate (at $x=\pm L_x/2$), characterized by a grain-wall
restitution coefficient $\alpha_w$, the inferior wall (at $y=0$) is a
Plexiglas\textsuperscript{\textregistered} piston vibrating with
amplitude $A$ and frequency $f$, giving an average squared velocity
$v_0^2=(A 2\pi f)^2/2$.  A high speed camera allowed us to reconstruct
the average fields $u(x,t)$, $n(x,t)$ and $T(x,t)$. Experiments have
been performed in a range of average 2D packing fraction
($\nu_{2D}=N\pi (d/2)^2/(L_xL_y)$) $\nu_{2D} \in [0.1 \%, 1 \%]$, with
a constant amplitude of vibration $A=1.85$ mm, a range of piston
oscillation frequency $f \in [10, 45]$ Hz and a range for the
inclination angle $\theta \in [0.011, 0.130]$ radiants, leading to a
range of rescaled maximum acceleration $\Gamma=A(2\pi f)^2/g_{eff} \in
[70,3800]$.  In all our experiments, with the only exception of the
lowest number of particles ($N=100$, where the mean free path is larger than the box's sides) we observed convection with two
convective cells that span the full width of the 2D plane. In all
convective cases density and temperature fields display gradients not
only in the $y$ but also in the $x$direction (i.e. transverse to
gravity): in particular, temperature decreases moving along the
horizontal from the lateral walls toward the center of the system, as
expected in view of the outgoing energy flux at the boundaries. Most
of the qualitative trends appear similar when $g$ or $N$ are
increased, suggesting that these two parameters play an analogous role. In particular
the centers of the convective cells move toward the lowest corners
(left or right) of the system. 

All these features have been confirmed, even in a larger range of
parameters, by MD numerical simulations: we have employed an
event-driven code for inelastic hard disks. Between collisions the
disks perform ``free'' ballistic fligths under the action of effective
gravity $g_{eff}$. When two disks collide, a normal instantaneous
collision with constant restitution coefficient $\alpha \in [0,1]$ is
performed. When a disk collides with the base wall, the velocity
component parallel to the base is left unaltered, while the normal
component is set to a new value extracted from a Maxwell-Boltzmann
distribution with temperature $v_0^2$ \cite{AG97}. When a disk collides
with the top wall, an elastic collision is performed. When a disk
collides with the lateral walls, an inelastic collision with an
infinite mass and with restitution coefficient $\alpha_w \in [0,1]$ is
performed. Some simulations with the presence of friction along the
ballistic flight (introduced by time-discretized dynamics) - which
would be analogous to the rolling/sliding friction between spheres and
the setup's floor - have also been performed and discussed in~\cite{noi},
giving very similar results. The simulations clearly show that
with the parameters of the experiment and the artificial choice of
elastic lateral walls the convection is totally suppressed, as
expected for experiments at low gravity~\cite{meerson1,meerson2}.  In
the simulations we have also shown that the intensity of convection
decreases linearly with $\alpha_w$, in a way similar to the
observation of~\cite{windows13} and that the width of the convective
cell decreases when $g_{eff}$ increases. When $g_{eff} \sim g$
convection is barely visible: this can explain the results and the
phase diagram described in~\cite{lohse07}. Interestingly, our
simulations allowed us to verify that DLW convection is present even
when grain-grain collisions are elastic ($\alpha=1$) provided that
$\alpha_w<1$: this suggests that the elastic limit is smooth (a good
hint for future theoretical research) and that DLW are sufficient to
create not only the horizontal gradient but also the vertical ones,
necessary for the developement of convection.
\section{New results}
\label{new}
\begin{figure}
  \includegraphics[width=7.5cm,height=5.7cm]{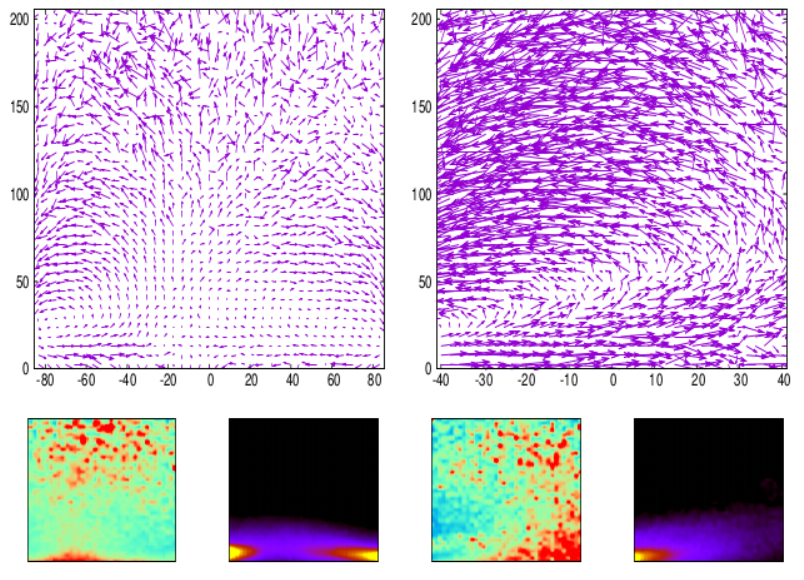}
\caption{ Fields from two experiments with $N=1000$, $f=45$ Hz,
  $A=1.85$ mm ($v_0=370$ mm/s) and (A): perpendicular lateral walls;
  (B): lateral walls inclined of $12$ degrees s (see sketch in the
  lower-right corner of Fig. 1). Top row: the velocity
  field with the two convective cells; coordinates are in units
  of particle's diameters ($=1$ mm). Bottom row: each couple of graphs
  show the (mass-free) temperature $T/m$ (left) and local packing
  fraction $\nu=n \pi (d/2)^2$ (right)  corresponding to the
  case in the top row. The $T$ scale goes from black (colder) at $T=0$
  to red (hotter) at $T=T_{max}$. The $\nu$ scale goes from black
  (more dilute) at $\nu=\nu_{min}$ to yellow (denser) at
  $\nu=\nu_{max}$. Values of $T_{max}/mv_0^2$ are: A) $0.08$, B) $0.15$.
  Ranges for ($\nu_{min},\nu_{max}$) are: A) $(0.05 \%, 0.5 \%)$, B)
  $(0.05 \%, 0.5\%)$.}
\label{fig:exp}
\end{figure}

\begin{figure}
    \includegraphics[width=7cm,height=5cm]{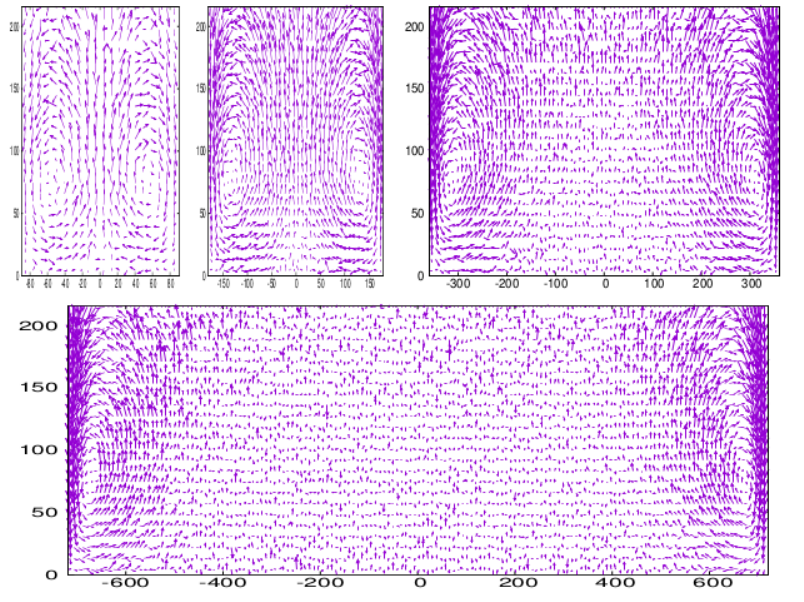}
    \caption{ Fields from four simulations with different width and
      same number of layers at rest ($N \sigma/L_x = 2.78$). From
      top-left: $L_x = 180$ ($N = 500$), $L_x = 360$ ($N = 1000$),
      $L_x = 720$ ($N = 2000$), $L_x = 1440$ ($N = 4000$).}
\label{fig:simw}
\end{figure}

\begin{figure}
    \includegraphics[width=7cm,height=7cm]{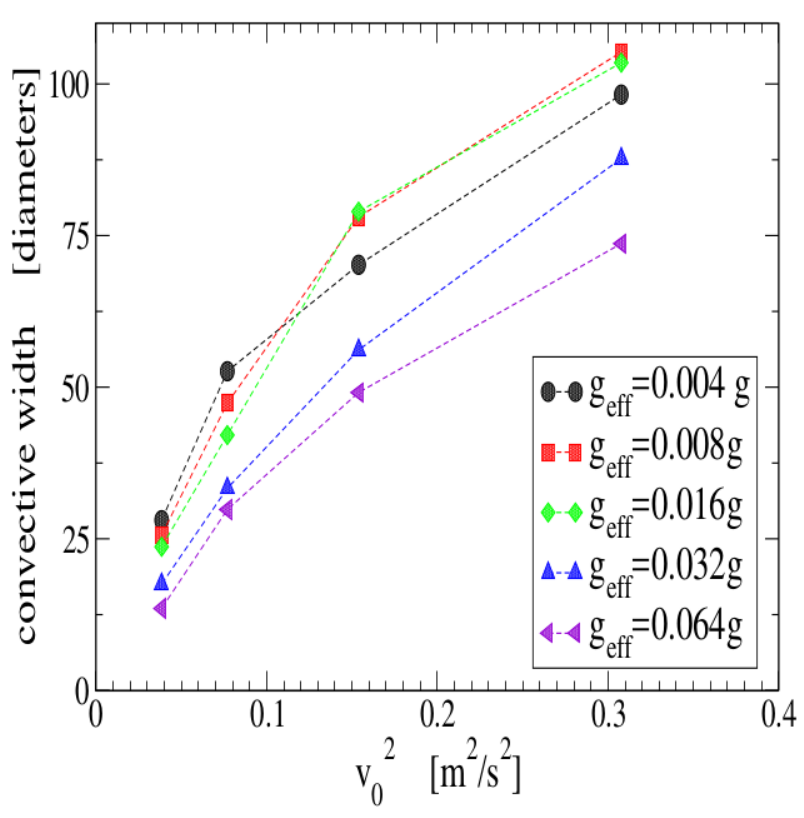}
    \caption{ Width of the convection roll ($L_x=720$
      and $N=2000$ particles) as a function of $v_0^2$ for different
      values of $g_{eff}$.}
\label{fig:simv0}
\end{figure}

In Fig.~\ref{fig:exp} we report the results of two different
experiments with the same number of particles ($N=1000$) and same
driving parameters, but different geometry of the box. The experiment
on the left is performed with the usual setup whereas on the right we
show a different set-up where the lateral walls form an inwards angle
of $12$ degrees with respect to the bottom wall direction. Apparently,
a small change of inclination of the walls leads to quite an important
qualitative change in the velocity, density and temperature field, in
particular here we observe only a single convective cell. We stress
that we have carefully checked the system to be symmetric with respect
to inversion of the $x$ direction, but it is possible that small
asymmetries escape our scrutiny. We have not a simple
  interpretation of this result, and in particular of the symmetry
  breaking. We expect that within long experiments both clockwise and
  anti-clockwise convections should be observed, separated by
  ``transition times'' which could be very long. We underline that the
  general mechanism for convection discussed after
  Eq.~\eqref{2d_balance} does not force the presence of two symmetric
  cells, it only states that a non-convective steady state is
  forbidden. In particular, it does not prevent an oscillation between
  single cells with opposite rotations.
In Fig.~\ref{fig:simw} we report results from different MD simulations
with the same number of layers at rest $N\sigma/L_x$ (characterising
the amount of mass per unit of horizontal length which is the
parameters replacing density in an isotropic system such as ours) but
different widths of the system, illustrating the remarkable fact that
the width of the cell (at fixed layers at rest, gravity and energy
driving conditions) is constant and does not depend of the width. This
is in perfect agreement with the idea that DLW-TC is a ``boundary
effect'', but nevertheless it spans a macroscopic portion of the
system, i.e. a large number of mean free path (see~\cite{noi} and its
Supplemental Materials for a discussion of Knudsen numbers in our
system).
In Fig.~\ref{fig:simv0} other numerical results allow us to
understand, for several values of $g_{eff}$, the important role of
$v_0^2$ which represents the intensity of vibration of the base
piston, i.e. the unique source of energy in the system. For all values
of $g_{eff}$ the width of the convective cell increase with $v_0^2$,
even if such a growth slows down at high $v_0^2$. Consistently with
previously reported data~\cite{noi}, the width of the cell decreases
monotonically with $g_{eff}$ for low values of $v_0$: on the contrary, it presents non-monotonicy with an initial growth at the first studied values of $g_{eff}$ followed by a decrease in $g_{eff}$, for larger values of $v_0$.
\section{Conclusions}
\label{conclusions}
In conclusion, we have discussed recent results on a novel thermal
(buoyancy-driven) convection effect induced by dissipative lateral
walls in vibrated granular gases, which at low gravity can be better
isolated from the usual convection induced by bulk (grain-grain)
inelasticity~\cite{noi}. In this manuscript we have presented further
evidences about the complexity of this phenomenon. In particular we
have shown that the inclination of the lateral walls affects the
convection in an unexpected way, and we have shown new numerical
results with systems of different widths, supporting the idea that
DLW-TC is a ``boundary-effect'', i.e. it is limited to a (macroscopic)
region close to the lateral walls. DLW-TC is replaced by usual thermal
convection when the parameters are chosen in the region of
instability~\cite{meerson1,meerson2}, e.g. for high enough bulk
inelasticity, as shown in the Supplemental Materials of~\cite{noi}. We
remark that while many studies in {\em dense} vibrated granular
materials have shown boundary-induced convection, and while other more
recent studies in vibrated granular gases have shown the existence of
an important role of the lateral wall, the study in~\cite{noi} is the
first to demonstrate the existence of a wall-induced convection
in gases which is well distinct from the bulk-inelasticity-induced
effect. Such a result is even more fascinating as a theoretical
explanation is lacking and cannot be retrieved in the form of a
stability analysis (as it was done for bulk thermal convection)
because a non-convective simpler base state (either homogeneous or
unidimensional \cite{VU09}) does not exist (i.e., the DLW convective
state is the base steady state). This, of course has further
implications, for instance, the stability threshold for the BT
convection will need to be calculated starting out from the DLWc state
and not from a non-convective one.
\section*{Acknowledgements}
F. V. R. acknowledges support from Grants FIS2016-76359-P (Spanish Gov.) and GRU10158 (Junta de Extremadura, partially financed by ERDF
funds).
%
%
\bibliography{biblio_merged}
\end{document}